**The correlated linking numbers of a Brownian loop with two arbitrary curves**

J.H. Hannay,
H.H. Wills Physics Laboratory,
University of Bristol,
Tyndall Ave,
Bristol BS6 7LP, UK.

*Abstract*
The standard kinetic path integral for all spatially closed Brownian paths (loops) of duration *t* weighted by the product *mn* is evaluated, where *m* and *n* are the linking numbers of the Brownian loop with two arbitrary curves in 3D space. The path integral thus indicates the extent to which these two linking numbers are correlated, ranging from the value zero for far apart curves when it is unlikely that the Brownian loop links with both, to ±infinity for nearly coincident curves.  The result takes a form that loosely resembles that for the mutual inductance of two current carrying circuits in magnetostatics, a double line integral, but is also dependent on a single extra parameter, the duration *t* of the path. The result for the equivalent two-dimensional problem was given previously [Hannay 2018].

*Introduction*
Here *I,* a rather natural path integral (1), that has both kinetic and topological ingredients, is to be evaluated.  The input is two arbitrarily chosen fixed curves in three dimensional space referred to as 'circuits' (the curves must have no ends – they must be loops or be infinitely long).  Besides the two circuits there is an additional input: a single positive parameter *t*.  The path integral *I* has the standard kinetic exponent for diffusion, or mathematical Brownian motion, of duration *t* in three dimensions. The paths have their initial and final positions coincident, and this position is integrated over all space (thus all 'Brownian loops' are covered).  The path integral is weighted by the product *mn* of the two linking numbers (2) of the loop with the pair of circuits (Fig 1).  It thus indicates the extent to which these two linking numbers are correlated, ranging from the value zero for far apart circuits when it is unlikely that the Brownian loop links with both, to ±infinity for nearly coincident circuits.  The equivalent problem in two dimensions is that of a Brownian loop's correlated winding numbers around two arbitrary points in a plane.  The corresponding path integral (the same as *I* with the 3s replaced by 2s) for was calculated recently [Hannay 2018].

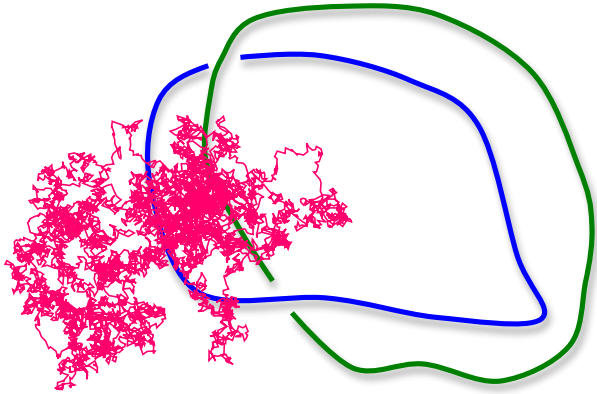

Fig 1. A Brownian loop in 3D may or may not link with either, or both, of two arbitrary circuits. Indeed it may wind arbitrarily many times, $m$ and $n$, around them (the linking numbers). The two circuits need not be loops as shown here, but could extend to infinity (e.g. straight lines). The path integral (1) over all Brownian loops weighted by the product $mn$ is evaluated (3). This (being non-zero) shows that the linking numbers are correlated, and loosely resembles the formula for the mutual inductance between the two circuits.

The evaluation of the path integral $I$ will lead to a double line integral formula (3) loosely resembling the mutual inductance of two electrical circuits in magnetostatics [Feynman 1964]. Indeed, as mentioned later, in the limit of large $t$, $I\sqrt{t}$ is directly proportional to their mutual inductance. As in the case mutual inductance, the two circuits may be separate or arbitrarily knotted or linked with each other (it is the linking of each individually with the Brownian loop that matters). The value of $I$ does not jump suddenly if the two circuits are deformed to change their topology by crossing through themselves or each other.

The motivation for evaluating $I$ comes mainly from its privileged simplicity – there seems to be no simpler topologically weighted free space path integral that is not zero or infinite or intractable. Linking path integrals of a Brownian loop with a single circuit (of general shape), instead of two circuits, are all: zero (e.g. weighting $m$), or infinite (e.g. weighting $m^2$), or intractable (e.g weighting $(1-\delta_{m0})$). The zero case derives from time reversal symmetry. The infinity case can be justified by the present calculation of $I$: if the two circuits involved are nearly coincident, nearly a single circuit, the weighting $mn$ in $I$ tends to $m^2$. In this limit $I$ diverges as discussed later, equation (17). One may speculate, incidentally, that all moments other than $mn$ that are not forced to be zero by time reversible symmetry, are infinite (e.g. $mn^3$), both in 2D and 3D.

Physical motivation is quite tentative at present. One context in which Brownian loops are approximately realised is that of closed polymer chains or of turbulent vortex loops, though it is not obvious what role the linking correlation could play. A more promising context is quantum statistical mechanics where taking the required trace of the density matrix at temperature $T$ is expressed, in terms of Feynman paths (or Brownian ones since the exponent is real instead of imaginary), as requiring that the paths be spatially closed after a duration $\hbar/kT$. An instance, admittedly somewhat contrived, is the joint susceptibility of a quantum charge in the presence of two Aharonov-Bohm magnetic flux tubes with fluxes $\Phi_M$ and $\Phi_N$. If $Z(\Phi_M, \Phi_N)$ denotes its quantum partition function at temperature $T$, then $Z(\Phi_M, \Phi_N) - Z(0,0)$ is finite, and $\partial^2[Z(\Phi_M,\Phi_N) - Z(0,0)]/\partial\Phi_M\partial\Phi_N$ at $\Phi_M = \Phi_N = 0$ is proportional to $I$ (with $t \to \hbar/kT$ as is standard). Speculatively, a less contrived physical application of $I$ in the quantum statistical context may prove possible.

Before beginning, one side point about normalization may be noted. For statistical purposes, some normalization may be sought; a denominator path integral with which to divide $I$ to supply an expectation value, for example a correlation. Fortunately not all applications of $I$ require this (the example just outlined does not), because there seems to be no form of the denominator in 3D, and only one in 2D, that is both tractable and non-infinite. For instance, a denominator that merely has $mn$ replaced by unity gives infinity (as discussed shortly). The tractable one in 2D is the correlation [Hannay 2018] of winding numbers $m$ and $n$ defined by the expectation of $mn$ subject to the condition that one of the winding numbers $m$ is non-zero. The condition is imposed by replacing the product $mn$ with $(1-\delta_{m0})$ in denominator path integral, which is tractable in 2D but not in 3D.

However, $I$ can be used for solving a modified statistical problem, an expectation value, though not a correlation, which works for both 2 and 3 dimensions. Returning to the proposal of a denominator path integral with $mn$ replaced by unity, this path integral is infinite (because of the integral over all coincident initial and final points) and the ratio $I/denominator$ is therefore trivially zero. But this triviality can be compensated if, not just one Brownian loop is admitted, but a number density $\sigma$ of independent loops throughout space. If the denominator volume integral (over coincident initial and final points) is restricted to a large finite volume $V$, then in the limit of large $V$, the combination $\sigma V I/denominator$ is the (finite) expectation value of the linking number product summed over all the Brownian loops.

The path integral $I$ to be evaluated has the standard kinetic exponent for diffusion, or mathematical Brownian motion, of duration $t$ in three dimensions. For notational simplicity the units of time are chosen such that $\langle \Delta x^2 \rangle = \langle \Delta y^2 \rangle = \langle \Delta z^2 \rangle = \Delta t$. The paths have their initial and final positions $\mathbf{r}_0$ coincident, and this position is integrated over all space (thus over all 'Brownian loops'). With dot denoting $d/d\tau$, and the path integration denoted by $\int ... \int$ integral, $I$ is defined by:

$$I \equiv \int d^3\mathbf{r}_0 \int_{\mathbf{r}(0)=\mathbf{r}_0}^{\mathbf{r}(t)=\mathbf{r}_0} ..... \int m[\mathbf{r}(\tau)]\, n[\mathbf{r}(\tau)] \exp\left(-\frac{1}{2t}\int_0^t \dot{\mathbf{r}}(\tau)^2\, d\tau\right) \frac{d^{3\infty}\mathbf{r}(\tau)}{\sqrt{2\pi t/\infty}^{3\infty}} \qquad (1)$$

where the integer valued functionals $m$ and $n$ are the linking numbers (2) of the path with the two specified circuits $M$ and $N$. (To make definite the meaning of the notation in (1) it may be helpful to state that the value of the right hand side with the symbols $m[\mathbf{r}(\tau)]$ and $n[\mathbf{r}(\tau)]$ and $\int d^3\mathbf{r}_0$ deleted would be $1/(2\pi t)^{3/2}$).

Let $M$ be described by the arbitrarily parametrized vector function **M**($s$) and let **r**($t$) (with 0<$\tau$ <$t$) describe the Brownian motion (defined, following Levi, [Ito and McKean 1965] as a polygon in 3D, with $2^j$ steps in the limit as $j \to \infty$). The standard Gauss linking number is defined, with **r** denoting **r**($t$) and **M** denoting **M**($s$), by

$$m[\mathbf{r}(\tau)] \equiv \frac{1}{4\pi} \int \int \frac{(\mathbf{r} - \mathbf{M}) \cdot (d\mathbf{r} \wedge d\mathbf{M})}{|\mathbf{r} - \mathbf{M}|^3} \qquad (2)$$

This is independent of the parametrization chosen for $M$ except that its sign is determined by the choice of the direction of increasing $s$ along $M$. Given this choice, the linking number has a standard geometrical interpretation. Projected, that is, viewed from infinity, in an arbitrary direction, the loop **r** necessarily appears to cross the circuit **M** an even number of times. If at a crossing, the tangent vector of the closer member would have to be rotated anticlockwise (through an angle less than $\pi$) to align with that of the further member, the crossing is counted as positive, otherwise negative. Half the sum of these signs is the linking number. An equivalent formula to (2) defines $n[\mathbf{r}(\tau)]$.

Previous literature on path linking, path integrals in general, and Brownian motion includes the following: [Edwards 1967], [Hannay 2010], [Hannay 2001], [Hannay 1995], [Kac 1966], [Kleinert 2006], [Comtet and Tourigny 2015], [Ito and McKean 1965].

*Calculation*
The result to be derived is, with $\rho$ denoting |**M**−**N**|,

$$I = \frac{1}{8\sqrt{2\pi^5 t}} \int\int_{M\ N} \left[ \exp\left(-\frac{2\rho^2}{t}\right) - \sqrt{\pi} \sqrt{\frac{2\rho^2}{t}} \, \text{erfc}\sqrt{\frac{2\rho^2}{t}} \right] \frac{d\mathbf{M} \cdot d\mathbf{N}}{\rho} \qquad (3)$$

This is reminiscent of the formula for the mutual inductance of two current carrying circuits, e.g. [Feynman, Leighton, Sands, 1964], for which the square bracket factor is absent, and the constant outside the integrals is $-\mu_0 / 4\pi$.

An intermediate result of importance is a double surface integral (5) for $I$, enroute to the double line integral of the result (3). In (5) one surface has its edge, or boundary on the circuit $M$, and the other on $N$. The two surfaces are each simply connected but are otherwise arbitrary. It is very convenient to continue to use the labels $M$ and $N$ and vectors **M** and **N**, but to let these refer to either the circuits *or the surfaces*, the distinction being indicated by the integration line element d**M**, or area element d²**M** (and similarly for $N$). Although logically one should first derive (5) and thence the result (3), the form of (5) has a certain simplicity as follows, but its derivation is somewhat laborious, and closely follows [Hannay 2018], so it seems preferable first to propose (5), and show how it leads to (3), and to relegate its derivation to the appendix.

If the symbol **0** is used for **r₀**, the initial (and final) position in 3D space, the set of all paths (Brownian motions) from **0** back to **0** in the path integral (1) is reduced in the appendix to a 2×2=4-dimensional set to be integrated over. These go from **0** straight to $d^2$**M** on the surface **M**, then straight to $d^2$**N** on the surface **N**, and finally straight back to **0**, so $\mathbf{0} \to \mathbf{M} \to \mathbf{N} \to \mathbf{0}$. The alternative ordering $\mathbf{0} \to \mathbf{N} \to \mathbf{M} \to \mathbf{0}$ needs to be included as well. Each of the three stages is associated with duration, the three summing to *t*, and a probability density *P* for its Brownian vector displacement in that duration. The definition of *P*, with **A** and **B** being its endpoints, and with $\tau^B > \tau^A$, is the spreading Gaussian solution of the diffusion equation $\frac{1}{2}\nabla^2 P = \partial P/\partial t$ (with the units of time chosen to absorb the diffusion coefficient),

$$P^{\mathbf{AB}} \equiv \frac{1}{\left(2\pi(\tau^B - \tau^A)\right)^{3/2}} \exp\left(\frac{-|\mathbf{A} - \mathbf{B}|^2}{2(\tau^B - \tau^A)}\right) \tag{4}$$

In the formula (5) for *I* there are four such terms each with the same chain of superscripts 0M MN N0 signifying the three stage path. The terms differ in their arrangement of the two subscript letters M and N. These subscripts signify directional derivatives along the normal direction to $d^2$**M** or $d^2$**N**. Different signs on the terms are also needed. Finally the timings of the two passages through **M** and **N** need to be integrated over, maintaining their order.

$$I = \int d^3\mathbf{0} \int d^2\mathbf{M} d^2\mathbf{N} \int_0^t d\tau^N \int_0^{\tau^N} d\tau^M$$
$$\frac{1}{4}\left[P^{\mathbf{0M}} P^{\mathbf{MN}}_{\mathbf{M}} P^{\mathbf{N0}}_{\mathbf{N}} - P^{\mathbf{0M}}_{\mathbf{M}} P^{\mathbf{MN}} P^{\mathbf{N0}}_{\mathbf{N}} - P^{\mathbf{0M}} P^{\mathbf{MN}}_{\mathbf{MN}} P^{\mathbf{N0}} + P^{\mathbf{0M}}_{\mathbf{M}} P^{\mathbf{MN}}_{\mathbf{N}} P^{\mathbf{N0}}\right] + \text{Same with } M \leftrightarrow N$$

$$(5)$$

The piece of the integral *I* (5) coming from '*Same with* $M \leftrightarrow N$' is equal to the rest, and so can be replaced with a factor 2 multiplying the rest. The integral over $d^3\mathbf{0}$, is a lengthy but straightforward Gaussian integral. With $\tau^{MN} \equiv \tau^N - \tau^M$ and $\tau^{NM} \equiv t - \tau^{MN}$, it generates the integrand of (6) apart from the numerator factor $\tau^{NM}$. This latter factor is introduced by a trivial time integration, noting that the integrand depends only on the interval $\tau^N - \tau^M$ and not the time at which this interval starts (the start time therefore has a free range of $\tau^{NM}$). One is left with:

$$I = 2\int_0^t d\tau^{MN} \int\int d^2\mathbf{M} d^2\mathbf{N} \frac{\tau^{NM}}{32\pi^3(\tau^{MN}\tau^{NM})^{7/2}} \exp\left(-\frac{(\mathbf{M}-\mathbf{N})^2 t}{2\tau^{MN}\tau^{NM}}\right) \times$$
$$\left[(\mathbf{M}-\mathbf{N})\cdot\boldsymbol{\mu}\,(\mathbf{M}-\mathbf{N})\cdot\boldsymbol{\upsilon}\,(\tau^{MN} - \tau^{NM})^2 - (\boldsymbol{\mu}\cdot\boldsymbol{\upsilon})\tau^{MN}\tau^{NM} t\right] \tag{6}$$

where the **μ** and **ν** are the unit vectors perpendicular to $d^2$**M** and $d^2$**N** (in the positive sense according with the chosen sense of the boundary circuit).

The remaining time integral can then be done noting $(t^{MN} - t^{NM})^2 = (t - 2t^{MN})^2 = t^2 - 4t^{MN}t^{NM}$, and using Feynman and Hibbs [1965], eqn A4, namely

$$\int_0^T \exp\left(\frac{ia}{T-\tau} + \frac{ib}{\tau}\right) \frac{(T-\tau)d\tau}{(\tau(T-\tau))^{3/2}} = \sqrt{\frac{i\pi}{bT}} \exp\left(\frac{i}{T}(\sqrt{a} + \sqrt{b})^2\right) \tag{7}$$

for *a* and *b* with non-negative imaginary parts (indeed for the present application they are purely positive imaginary numbers). Setting *a=b* and then differentiating once and twice with respect to *a* yields the needed forms. The result, with $\boldsymbol{\rho} \equiv \mathbf{N} - \mathbf{M}$, and $\rho \equiv |\boldsymbol{\rho}|$ and $\hat{\boldsymbol{\rho}} \equiv \boldsymbol{\rho}/\rho$, is

$$I = \iint d^2\mathbf{M} d^2\mathbf{N} \exp\left(-\frac{2\rho^2}{t}\right) \frac{1}{\pi^{5/2} 4\sqrt{2t}\rho^3} \times \\ \left[\left(\frac{2\rho^2}{t} + \frac{3}{2}\right)\hat{\boldsymbol{\rho}} \cdot \boldsymbol{\mu} \, \hat{\boldsymbol{\rho}} \cdot \boldsymbol{\upsilon} - \boldsymbol{\mu} \cdot \boldsymbol{\upsilon}\left(\frac{2\rho^2}{t} + \frac{1}{2}\right)\right] \tag{8}$$

To demonstrate, via Stokes' theorem, that this double surface integral equals the desired double line integral along the circuits (3) it seems easiest to work backwards. Let the multiplier of d**M**•d**N** in (3), which is a scalar function of $\rho \equiv |\boldsymbol{\rho}| \equiv |\mathbf{N} - \mathbf{M}|$, be strategically written as the function $\int_\rho^\infty \tilde{\rho}\lambda(\tilde{\rho})d\tilde{\rho}$ where

$$\lambda(\rho) \equiv \frac{1}{2^{7/2}\pi^{5/2}\sqrt{t}\rho^3} \exp\left(-\frac{2\rho^2}{t}\right) \tag{9}$$

Then denoting $d\lambda/d\rho$ by $\lambda'$,

$$\iint\left(\int_\rho^\infty \tilde{\rho}\lambda(\tilde{\rho})d\tilde{\rho}\right)d\mathbf{M} \cdot d\mathbf{N} = \int \int \nabla_N \wedge \left(\int_\rho^\infty \tilde{\rho}\lambda d\tilde{\rho} \, d\mathbf{M}\right) \cdot \boldsymbol{\upsilon} \, d^2\mathbf{N} \tag{10}$$

$$= \oint \int \rho\lambda \, (d\mathbf{M} \wedge \hat{\boldsymbol{\rho}}) \cdot \boldsymbol{\upsilon} \, d^2\mathbf{N} = \int d^2\mathbf{N} \oint \rho\lambda \, (\hat{\boldsymbol{\rho}} \wedge \boldsymbol{\upsilon}) \cdot d\mathbf{M} \tag{11}$$

$$= \int d^2\mathbf{N} \int d^2\mathbf{M}\left[\nabla_M \wedge (\rho\lambda \, \hat{\boldsymbol{\rho}} \wedge \boldsymbol{\upsilon})\right] \cdot \boldsymbol{\mu} \tag{12}$$

$$= \int d^2\mathbf{N} \int d^2\mathbf{M}\left[(\boldsymbol{\upsilon} \cdot \nabla_M)(\rho\lambda\hat{\boldsymbol{\rho}}) - \boldsymbol{\upsilon}\nabla_M \cdot (\rho\lambda\hat{\boldsymbol{\rho}})\right] \cdot \boldsymbol{\mu} \tag{13}$$

$$= \int d^2\mathbf{N} \int d^2\mathbf{M}\left[\boldsymbol{\upsilon}(\rho\lambda' + 3\lambda) - \lambda'(\boldsymbol{\upsilon} \cdot \hat{\boldsymbol{\rho}})\boldsymbol{\rho} - \lambda\boldsymbol{\upsilon}\right] \cdot \boldsymbol{\mu} \tag{14}$$

$$= \int d^2\mathbf{N} \int d^2\mathbf{M}\left[(\boldsymbol{\upsilon} \cdot \boldsymbol{\mu})(\rho\lambda' + 2\lambda) - \rho\lambda'(\boldsymbol{\upsilon} \cdot \hat{\boldsymbol{\rho}})(\boldsymbol{\mu} \cdot \hat{\boldsymbol{\rho}})\right] \tag{15}$$

which equals the double surface integral (8) as required.

*Limits*

Three limiting forms of the result $I$ (1) can be identified, remembering that the units of time (having absorbed the diffusion coefficient) are such that square root of the duration $t$ has the dimensions of length.

(i) The easiest limit applies if $\sqrt{t}$ is much longer than the longest separation $\rho$ in the integrations in $I$ (3). (This can obviously only be the case if the two circuits are genuine closed curves not infinite). Then the square bracket in (3) is unity, and $I\sqrt{t}$ is directly proportional to the mutual inductance between the two circuits.

(ii) At the opposite extreme $\sqrt{t}$ is much shorter than the shortest separation $\rho$ in the integrations in $I$ (3). Generically there will be a single pair of points, one on each circuit, for which their separation $\rho$ has its minimal value $L$, say. The integrals can be performed in the saddle point approximation based on this pair of points. The tangent vectors at those points are orthogonal to the separation vector, and make an angle $\theta$, say, with each other. If $\gamma$ denotes the argument of the erfc in (3), the square bracket tends to $\exp(-\gamma^2)/2\gamma^2$ for small $t$. With $s^M$ and $s^N$ as displacements from the positions of minimal separation along the circuits, one has, with

$$\rho^2 = L^2 + (s^M)^2 + (s^N)^2 - 2s^M s^N \cos\theta,$$

$$I \approx \frac{1}{8\sqrt{2\pi^5 t}}\iint \exp\left(-\frac{2\rho^2}{t}\right)\frac{t\cos\theta\, ds^M ds^M}{4\rho^3} \approx \frac{1}{2\sqrt{2\pi^5 t}}\exp\left(-\frac{2L^2}{t}\right)\frac{\pi t^2}{8L^3 \tan\theta} \tag{16}$$

The divergence if $\theta=0$ would signify the breakdown of this simple saddle point approximation for the small $t$ limit.

(iii) An intermediate case for $\sqrt{t}$ has a different quantity involved in the limit, corresponding to coalescence of the two circuits onto a single circuit. In the limit of small separation the two circuits can be considered as the two edges of a very narrow ribbon whose width $w(s)$ is allowed to vary along the ribbon. The length $\sqrt{t}$ is supposed much greater than the (maximum) width, but much smaller than the length of the ribbon (indeed small enough that the Brownian loop always 'sees' a locally straight ribbon). Then with $s$ measuring distance along the ribbon centre line, the second term in (3) is negligible compared to the first and

$$I \approx \frac{1}{8\sqrt{2\pi^5 t}} \oint ds \int_{-\infty}^{\infty} \exp\left(-\frac{2(\Delta s^2 + w(s)^2)}{t}\right) \frac{d\Delta s}{\sqrt{\Delta s^2 + w(s)^2}}$$

$$= \frac{1}{16\sqrt{\pi^5}} \oint ds \, \exp\left(-\frac{w(s)^2}{t}\right) K_0\left(\frac{w(s)^2}{t}\right) \approx -\frac{1}{4\sqrt{\pi^5}} \oint ds \left(\gamma + \log\frac{w(s)^2}{2t}\right) \quad (17)$$

where $\gamma$ is Euler's Gamma. Thus $I$ is logarithmically divergent for close circuits, as is the mutual inductance of close conducting circuits in magnetostatics – the self inductance of a single circuit is infinite. (A formula for the limit of mutual inductance in this case is more awkward however, because the exponential cut off factor for large $\Delta s$ in (17) is lacking).

*Circular hoops example*
Returning to the exact formula (3) for $I$, a simple example allows the connection to be made between the present result (3) and that obtained earlier [Hannay 2018] for the windings correlation of Brownian loops in two dimensions. Suppose the two circuits are circular hoops of radius $R$ mutually displaced by a distance $L$ along their common axis. If the azimuthal angular separation of the integration elements along the hoops is $\Delta\phi$, the integral $I$ (3) is given, with
$\rho = \sqrt{L^2 + 4R^2 \sin^2(\Delta\phi/2)}$, by the single integral

$$I = \frac{1}{8\sqrt{2\pi^5 t}} \int_0^{2\pi} \left[\exp\left(-\frac{2\rho^2}{t}\right) - \sqrt{\pi}\sqrt{\frac{2\rho^2}{t}} \, \text{erfc}\sqrt{\frac{2\rho^2}{t}}\right] \frac{2\pi R^2 \cos\Delta\phi \, d\Delta\phi}{\rho} \quad (18)$$

It is the limit of this as $R$ tends to infinity, so that the circuits are locally parallel straight lines, that connects with the 2D analysis since projection along the lines essentially reduces the problem from 3D to 2D (a projected 3D Brownian motion being a 2D Brownian motion). No restriction is placed on the value of the duration $t$ (unlike the case (iii) in the previous section). In the $R \to \infty$ limit one can take $\cos\Delta\phi = 1$ and $\rho = \sqrt{L^2 + R^2\Delta\phi^2}$. With $\Delta s = R\Delta\phi$:

$$I \xrightarrow{R \to \infty}$$

$$\frac{2\pi R}{8\sqrt{2\pi^5 t}} \int_{-\infty}^{\infty} \left[\exp\left(-\frac{2(L^2 + \Delta s^2)}{t}\right) - \sqrt{\pi}\sqrt{\frac{2(L^2 + \Delta s^2)}{t}} \, \text{erfc}\sqrt{\frac{2(L^2 + \Delta s^2)}{t}}\right] \frac{d\Delta s}{\sqrt{L^2 + \Delta s^2}}$$

(19)

This is actually equal to $2\pi R/\sqrt{2\pi t}$ times the equivalent result (20) for $I$ in two dimensions [Hannay 2018] (that is, the evaluation of (1), but with the 3 replaced by 2, and $m$ and $n$ counting windings around two points, separation $L$ in a plane), namely

$$\frac{1}{8\pi^2}\int_0^1 d\beta\,\frac{1}{\beta}\times$$

$$\left(\exp\left[-\frac{1}{2\beta(1-\beta)}\frac{L^2}{t}\right]-\sqrt{\pi}\sqrt{\frac{1}{2\beta(1-\beta)}\frac{L^2}{t}}\mathrm{erfc}\left[\sqrt{\frac{1}{2\beta(1-\beta)}\frac{L^2}{t}}\right]\right) \quad (20)$$

The reconciliation of the integrals in (19) and (20), follows by mapping $b=\beta-\tfrac{1}{2}$ onto $s/L$ via the matching of the exponents: $[1+(s/L)^2](1-4b^2)=1$. The integrands are both even functions of their respective variables $b$ and $s$. The significance of the factor $2\pi R/\sqrt{2\pi t}$ just mentioned is this: the closure of a projected Brownian motion (forming a 2D loop) does not imply the closure of the 3D Brownian motion from which it was projected, since the two endpoints only need to align along the projection direction. Given such alignment, the probability density for actually being coincident in 3D is $1/\sqrt{2\pi t}$.

*Appendix: Derivation of the double area integral (5) for I.*
The justification of the double integral formula (5) for the linking number product closely follows that in the equivalent two dimensional problem [Hannay 2018]. With the present path integral notation (less emphasized before) it can be explained more economically as follows. Consider the quantities, or generating functions:

$$Q^{(M)0\bar{0}}\equiv \int_{\mathbf{r}(0)=\mathbf{0}}^{\mathbf{r}(t)=\bar{\mathbf{0}}}\!\!\!\!\!\dots\int m[\mathbf{r}(\tau)]\,\exp\!\left(-\frac{1}{2t}\int_0^t \dot{\mathbf{r}}(\tau)^2\,d\tau\right)\frac{d^{3\infty}\mathbf{r}(\tau)}{\sqrt{2\pi t/\infty}^{3\infty}} \quad (A1)$$

$$Q^{(N)0\bar{0}}\equiv \int_{\mathbf{r}(0)=\mathbf{0}}^{\mathbf{r}(t)=\bar{\mathbf{0}}}\!\!\!\!\!\dots\int n[\mathbf{r}(\tau)]\,\exp\!\left(-\frac{1}{2t}\int_0^t \dot{\mathbf{r}}(\tau)^2\,d\tau\right)\frac{d^{3\infty}\mathbf{r}(\tau)}{\sqrt{2\pi t/\infty}^{3\infty}} \quad (A2)$$

$$Q^{(MN)0\bar{0}}\equiv Q^{(NM)0\bar{0}}\equiv \int_{\mathbf{r}(0)=\mathbf{0}}^{\mathbf{r}(t)=\bar{\mathbf{0}}}\!\!\!\!\!\dots\int m[\mathbf{r}(\tau)]\,n[\mathbf{r}(\tau)]\exp\!\left(-\frac{1}{2t}\int_0^t \dot{\mathbf{r}}(\tau)^2\,d\tau\right)\frac{d^{3\infty}\mathbf{r}(\tau)}{\sqrt{2\pi t/\infty}^{3\infty}}$$
$$(A3)$$

The last of these quantities $Q$ (A18) is the same formula as $I$ in (1) except that the initial and final points are not required to be the spatially coincident and are not integrated over. The meanings of $m$ and $n$ are the signed count of the path $\mathbf{r}(\tau)$ crossing through the surfaces $M$ and $N$ respectively (thus correctly counting the windings in the special case of loop paths with coincident initial and final points). The first quantity $Q$ (A16) does not depend on the surface $N$ or its boundary circuit at all; $N$ can be considered absent, leaving only the surface $M$.

As a function of their final endpoints $\bar{\mathbf{0}}$ all three quantities obey the diffusion equation $\tfrac{1}{2}\nabla^2 Q = \partial Q/\partial t$ at all locations other than the surfaces named in their bracketed superscript labels – this follows from the fact that the counting numbers *m* and *n* change only on these surfaces. Across them there is a jump in the value of *Q* that is easily read off since the relevant counting numbers in the integrand change by unity as the final endpoint $\overline{\mathbf{0}}$ is displaced across the surface.

|  | As $\overline{\mathbf{0}}$ is moved across *M* | As $\overline{\mathbf{0}}$ is moved across *N* |
|---|---|---|
| Jump of $Q^{(M)0\overline{0}}$ = | $P^{0M}$ | 0 |
| Jump of $Q^{(N)0\overline{0}}$ = | 0 | $P^{0N}$ |
| Jump of $Q^{(MN)0\overline{0}}$ = | $Q^{(N)0\overline{0}}$ | $Q^{(M)0\overline{0}}$ |

Here *P* means the spreading Gaussian (5). In addition to these jumps there are also jumps in the corresponding normal derivatives, thus for example, Jump of $Q_M^{(M)0\overline{0}}$ as $\overline{\mathbf{0}}$ is moved across *M* is $P_M^{0\overline{0}}$, and Jump of $Q_M^{(MN)0\overline{0}}$ as $\overline{\mathbf{0}}$ is moved across *M* is $Q_M^{(N)0\overline{0}}$. These results are to be used in conjunction with a separate ingredient governing functions obeying the diffusion equation, namely Green's theorem.

Green's theorem will allow the value of a function obeying the diffusion equation in an arbitrary empty region of space to be expressed in terms of all the prior values (0<τ<*t*) and normal derivatives of the function on the *boundary* of the empty region (together with *initial* values, which in the present context will be zero since each *Q* is zero for zero duration paths *t*=0). To be empty, the region must exclude the surfaces *M* and *N*, so the boundary needs to be chosen as a sheath, or enveloping surface, fully enclosing *M* and another fully enclosing *N*, with the empty region outside the sheath. To be useful here the sheaths should wrap closely, so that each element of *M* and *N* has an element of sheath surface immediately on either side.

The relevant version of Green's theorem runs as follows. Let scalar fields *A* and *B* be solutions respectively of the diffusion equation and its time reverse in 3+1D spacetime (*x,y,z,t*), that is, $\frac{1}{2}A_{xx} + \frac{1}{2}A_{yy} + \frac{1}{2}A_{zz} = A_t$ and $\frac{1}{2}B_{xx} + \frac{1}{2}B_{yy} + \frac{1}{2}B_{zz} = -B_t$. Then in terms of differential forms, using these diffusion equations, it follows that a certain exterior derivative ($\tilde{d}$) vanishes:

$$0 = \tilde{d}\bigl[-AB(\tilde{d}x \wedge \tilde{d}y \wedge \tilde{d}z) + \tfrac{1}{2}(AB_x - BA_x)(\tilde{d}y \wedge \tilde{d}z \wedge \tilde{d}t)$$
$$+ \tfrac{1}{2}(AB_y - BA_y)(\tilde{d}z \wedge \tilde{d}x \wedge \tilde{d}t) \qquad (A4)$$
$$+ \tfrac{1}{2}(AB_z - BA_z)(\tilde{d}x \wedge \tilde{d}y \wedge \tilde{d}t)\bigr]$$

Consequently the integral of the square bracket 3-form over the boundary of an empty spacetime region is zero. In the circumstances relevant for the present problem, the scalar field *A* represents any one of the three quantities *Q*, and *B* represents the ordinary Gaussian diffusive sink field, the time reverse of the spreading Gaussian *P*, collapsing at a chosen time to a delta function at a target position $\bar{0}$. The first, purely spatial 3-form term, in (A4) supplies initial and final data, and the other three terms supply boundary data for all times in between. The $Q$ quantities ($A$) are zero initially, and $B$ is zero finally except at the target position $\bar{0}$, picking this point out in the $A$ that it multiplies. The left hand sides of the remaining equations in this appendix come from the first term of (A4) (with $B$ as the delta function), and the right hand sides come from the last three terms of (A4).

In order to find $Q^{(M)0\bar{0}}$ a single application of Green's theorem suffices, and similarly for $Q^{(N)0\bar{0}}$, whereas for the quantity $Q^{0MN\bar{0}}$ that is ultimately needed, two successive applications will be required.

Starting with $Q^{(M)0\bar{0}}$, its value for a general position of $\bar{0}$ at time $t$ is given by Green's theorem as

$$Q^{(M)0\bar{0}} = \int_0^t d\tau \int d^2\mathbf{M} \tfrac{1}{2}(Q^{(M)0M+}P_M^{M\bar{0}} - Q_M^{(M)0M+}P^{M\bar{0}})$$
$$+ \int_0^t d\tau \int d^2\mathbf{M}(-1)\tfrac{1}{2}[(Q^{(M)0M+} - P^{0M})P_M^{M\bar{0}} - (Q_M^{(M)0M+} - P_M^{0M})P^{M\bar{0}}] \qquad (A5)$$

$$= \int_0^t d\tau \int d^2\mathbf{M} \tfrac{1}{2}(P^{0M}P_M^{M\bar{0}} - P_M^{0M}P^{M\bar{0}}) \qquad (A6)$$

The front and back (sheath) surface integrals sandwiching $M$ are separated in (A19), with $M+$ denoting the front surface (again, the 'front' is fixed by the sense chosen for the boundary circuit). The back surface integral makes use of the jump given in the table above; the minus sign in front of the square bracket arises because the back surface outward normal is opposite in direction to the front one. Because $P^{0M}$ in (A5) is smooth (has no jump) across $M$, it is not necessary to distinguish between the front and back sheath surfaces with a subscript on $M$. The same applies to $P^{M\bar{0}}$, and to the normal derivatives of both. The formula for $Q^{(N)0\bar{0}}$ is obviously the same with $M$ replaced by $N$.

The desired quantity $Q^{(MN)0\bar{0}}$ at a general position of $\bar{0}$ at time $t$ is similarly expressed in terms of $Q^{(M)0\bar{0}}$ and $Q^{(N)0\bar{0}}$. Again using the jump table:

$$Q^{(MN)0\bar{0}} = \int_0^t d\tau^N \int_0^{\tau^N} d\tau^M \int d^2\mathbf{N} \int d^2\mathbf{M} \Big( \tfrac{1}{2}(Q^{(MN)0N+}P_N^{N\bar{0}} - Q_N^{(MN)0N+}P^{N\bar{0}})$$
$$+ (-1)\tfrac{1}{2}[(Q^{(MN)0N+} - Q^{(M)0N})P_N^{N\bar{0}} - (Q_N^{(MN)0N+} - Q_N^{(M)0N})P^{N\bar{0}}]\Big) + \text{ Same with } M \leftrightarrow N$$
$$(A7)$$

$$= \int_0^t d\tau^N \int_0^{\tau^N} d\tau^M \int d^2\mathbf{N} \int d^2\mathbf{M} \; \tfrac{1}{2}(Q^{(M)0N} P_N^{N\bar{0}} - Q_N^{(M)0N} P^{N\bar{0}})] \; + \; \text{Same with } M \leftrightarrow N$$

(A8)

$$= \int_0^t d\tau^N \int_0^{\tau^N} d\tau^M \int d^2\mathbf{N} \int d^2\mathbf{M} \frac{1}{4}\left[P^{0M} P_M^{MN} P_N^{N\bar{0}} - P_M^{0M} P^{MN} P_N^{N\bar{0}} - P^{0M} P_{MN}^{MN} P^{N\bar{0}} + P_M^{0M} P_N^{MN} P^{N\bar{0}}\right]$$

$$+ \; \textit{Same with } M \leftrightarrow N$$

(A9)

The forms from (A6) for the *Q*s in (A8) have been used to obtain (A9). Taking $\mathbf{0} = \bar{\mathbf{0}}$ and integrating over **0** yields the double surface form of *I* (5). This therefore completes the derivation of the main result (3), since the connection between the line and surface integrals was established in the main text.

*References*

Comtet, A. and Tourigny, Y., 2015. Explicit formulae in probability and in statistical physics. In *In Memoriam Marc Yor-Séminaire de Probabilités XLVII* (pp. 505-519). Springer, Cham. (arXiv Math 1409.3777)

Edwards, S.F., 1967. Statistical mechanics with topological constraints: I. *Proceedings of the Physical Society*, *91*(3), p.513.

Feynman RP and Hibbs AR 1965, *Quantum mechanics and Path integrals*, McGraw-Hill, New York.

Feynman RP, Leighton RB, and Sands M, 1964, *The Feynman lectures on physics*, vol 2, eqn 17.30. Addison-Wesley, Reading, Massachusetts.

Hannay, J.H., 2018. (with corrigendum) Winding number correlation for a Brownian loop in a plane. *Journal of Physics A: Mathematical and Theoretical*.

Hannay, J.H., 2010. Kirchhoff diffraction optics and the nascent Aharonov–Bohm effect: a theorem. *Journal of Physics A: Mathematical and Theoretical*, *43*(35), p.354001.

Hannay, J.H., 2001. Path–linking interpretation of Kirchhoff diffraction: a summary. *Philosophical Transactions of the Royal Society of London A: Mathematical, Physical and Engineering Sciences*, *359*(1784), pp.1473-1478.

Hannay, J.H., 1995. Path-linking interpretation of Kirchhoff diffraction. *Proc. R. Soc. Lond. A*, *450*(1938), pp.51-65.

Ito, K., and McKean H.P., 1965. *Diffusion processes and their sample paths.* (Springer-Verlag)